\documentclass[twocolumn,prl,aps]{revtex4}

\usepackage{epsfig}
\usepackage{axodraw}

\newcommand{\bed}{\begin{displaymath}}
\newcommand{\eed}{\end{displaymath}}
\newcommand{\beq}{\begin{equation}}
\newcommand{\eeq}{\end{equation}}
\newcommand{\bea}{\begin{eqnarray}}
\newcommand{\eea}{\end{eqnarray}}
\newcommand{\tgb}{{\rm tg}\beta}
\newcommand{\tga}{{\rm tg}\alpha}
\newcommand{\stgb}{{\rm tg}^2\beta}

\newcommand{\MS}{\overline{\rm MS}}
\newcommand{\sq}{\tilde{b}}
\newcommand{\st}{\tilde{t}}
\newcommand{\sgl}{\tilde{g}}
\newcommand{\Dmb}{\ensuremath{\Delta m_b}}
\newcommand{\lsim}{\raisebox{-0.13cm}{~\shortstack{$<$ \\[-0.07cm] $\sim$}}~}

\begin{document}

\title{
\begin{flushright} {\rm PSI--PR--08--12} \\ \end{flushright}
MSSM Higgs Couplings to Bottom Quarks: Two-Loop Corrections}

\author{David Noth}
\affiliation{Paul Scherrer Institut, CH-5232 Villigen PSI, Switzerland \\
and \\ Institut f\"ur Theoretische Physik, Z\"urich University,
CH-8057 Z\"urich, Switzerland}

\author{Michael Spira}
\affiliation{Paul Scherrer Institut, CH-5232 Villigen PSI, Switzerland}


\begin{abstract} 
\noindent
We present the two-loop SUSY-QCD corrections to the effective bottom
Yukawa couplings within the minimal supersymmetric extension of the
Standard Model. The effective Yukawa couplings include the resummation
of the non-decoupling corrections $\Delta m_b$ for large values of $\rm
tg\beta$. We have derived the two-loop SUSY-QCD corrections to the
leading SUSY-QCD and top-induced SUSY-electroweak contributions to
$\Delta m_b$.  The scale dependence of the resummed Yukawa couplings is
reduced from ${\cal O}(10\%)$ to the per-cent level. These results
reduce the theoretical uncertainties of the MSSM Higgs branching ratios
to the accuracy which can be achieved at a future linear $e^+e^-$
collider.
\end{abstract}


\maketitle

\noindent
The Higgs mechanism~\cite{Higgs:1964ia} is a cornerstone of the
Standard Model (SM) and its supersymmetric extensions. The masses of
the fundamental particles, electroweak gauge bosons, leptons, and
quarks, are generated by interactions with Higgs fields. The search
for Higgs bosons is thus one of the most important endeavors in
high-energy physics and is being pursued at the upgraded
proton--antiproton collider Tevatron~\cite{Carena:2000yx} with a
centre-of-mass (CM) energy of $1.96$~TeV, followed in the near future
by the proton--proton collider LHC~\cite{atlas_cms_tdrs} with $14$~TeV
CM energy.

The minimal supersymmetric extension of the Standard Model (MSSM)
requires the introduction of two Higgs doublets. After electroweak
symmetry breaking there are five elementary Higgs particles, two CP-even
($h,H$), one CP-odd ($A$) and two charged ones ($H^\pm$). At lowest
order all couplings and masses of the MSSM Higgs sector are fixed by two
independent input parameters, which are generally chosen as
$\tgb=v_2/v_1$, the ratio of the two vacuum expectation values
$v_{1,2}$, and the pseudoscalar Higgs mass $M_A$.  Including the
one-loop and dominant two-loop corrections the upper bound on the light
scalar Higgs mass is $M_h\lsim 135$ GeV \cite{mssmrad}. The couplings of
the various Higgs bosons to fermions and gauge bosons depend on mixing
angles $\alpha$ and $\beta$, which are defined by diagonalizing the
neutral and charged Higgs mass matrices.

The negative direct searches at LEP2 yield lower bounds of $M_{h,H} >
92.8$ GeV and $M_A > 93.4$ GeV. The range $0.7 < \tgb < 2.0$ in the MSSM
is excluded by the Higgs searches for a SUSY scale $M_{SUSY}=1$ TeV at
the LEP2 experiments \cite{lep2}.

The dominant genuine SUSY-QCD and SUSY-electro\-weak corrections to
bottom-Yukawa-coupling induced processes, as e.g.~Higgs boson decays to
$b\bar b$ pairs and Higgs radiation off bottom quarks, can be derived
from the effective Lagrangian \cite{Hall:1993gn,CGNW,GHS}
\bea
{\cal L}_{eff} & = & -\frac{m_b}{v} \bar b \left[ \tilde g_b^h h
+ \tilde g_b^H H - \tilde g_b^A i \gamma_5 A \right] b
\label{eq:leff}
\eea
with the resummed Yukawa coupling factors
\bea
\tilde g_b^h & = & \frac{g_b^h}{1+\Delta m_b} \left(
1-\frac{\Delta m_b}{\tga~\tgb}\right) \nonumber \\
\tilde g_b^H & = & \frac{g_b^H}{1+\Delta m_b} \left(
1+\Delta m_b \frac{\tga}{\tgb}\right) \nonumber \\
\tilde g_b^A & = & \frac{g_b^A}{1+\Delta m_b} \left(
1-\frac{\Delta m_b}{\stgb}\right)
\label{eq:respar}
\eea
where $\Dmb$ determines the relative corrections to the bottom Yukawa
couplings. The Higgs couplings are given by
\bea
g_b^h = -\frac{\sin\alpha}{\cos\beta}, \quad g_b^H =
\frac{\cos\alpha}{\cos\beta}, \quad g_b^A = \tgb
\eea
The leading one-loop corrections $\Dmb$ to these effective couplings can
be cast into the form
\bea
\Delta m_b & = & \Delta m_b^{QCD (1)} + \Delta m_b^{elw (1)} \nonumber \\
\Delta m_b^{QCD (1)} & = &
\frac{2}{3}~\frac{\alpha_s(\mu_R)}{\pi}~m_{\sgl}~\mu~\tgb~
I(m^2_{\sq_1},m^2_{\sq_2},m^2_{\sgl}) \nonumber \\
\Delta m_b^{elw (1)} & = & \frac{\lambda_t^2(\mu_R)}{(4\pi)^2}~A_t~\mu~\tgb~
I(m^2_{\st_1},m^2_{\st_2},\mu^2)
\label{eq:dmb}
\eea
with the scalar function
\bed
I(a,b,c) = \frac{\displaystyle ab\log\frac{a}{b} + bc\log\frac{b}{c}
+ ca\log\frac{c}{a}}{(a-b)(b-c)(a-c)}
\eed

The parameter $v=\sqrt{v_1^2+v_2^2} = \sqrt{1/{\sqrt{2}G_F}}$ is related
to the Fermi constant $G_F$ and $\lambda_t=\sqrt{2}m_t/v_2$ denotes the
top Yukawa coupling. The SUSY--QCD and top-induced SUSY--electroweak
corrections turn out to be significant for large values of $\tgb$ and
moderate or large $\mu$ and $A_t$ values. In order to improve the
perturbative result all terms of ${\cal
O}\left[(\alpha_s\,\mu\,\tgb/M_{SUSY})^n\right]$ and ${\cal
O}\left[(\lambda_t^2\,A_t\,\tgb/M_{SUSY})^n\right]$ have been resummed
in Eq.~(\ref{eq:respar}) \cite{CGNW,GHS}.  The correction \Dmb\ is
\textit{non-decoupling} in the sense that scaling \textit{all} SUSY
parameters $m_{\sq_{1,2}}, m_{\sgl}, \mu$ in Eq.~(\ref{eq:respar})
leaves \Dmb\ invariant. However, its contribution develops decoupling
properties \cite{decoup}. The corrections $\Dmb$ contain the strong
coupling $\alpha_s(\mu_R)$ and the top Yukawa coupling $\lambda_t
(\mu_R)$ with significant renormalization scale dependences. This leads
to theoretical uncertainties in e.g.~the MSSM Higgs boson decay widths
and branching ratios of up to ${\cal O}(10\%)$ \cite{GHS} which are
larger than the achievable accuracy at a future $e^+e^-$ linear collider
(ILC) \cite{ilc}. In this letter we present the two-loop SUSY--QCD
corrections to the contributions $\Dmb$ of Eq.~(\ref{eq:dmb}) in order
to reduce the theoretical uncertainties to the per-cent level.
\begin{figure}[htbp]
\vspace*{-0.3cm}

\epsfig{file=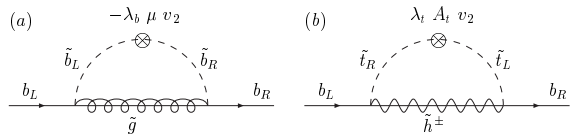,%
        bbllx=30pt,bblly=550pt,bburx=520pt,bbury=650pt,%
        scale=0.5,clip=}
\vspace*{-0.7cm}

\caption{\it \label{fg:one-loop} One-loop diagrams of (a) the SUSY-QCD
and (b) the top-induced SUSY-electroweak contributions to the bottom
self-energy with mass insertions corresponding to the
corrections $\Dmb$ of the bottom Yukawa couplings. The virtual particles
involve bottom quarks $b$, sbottom $\sq$ and stop $\st$ squarks, gluinos
$\sgl$ and charged Higgsinos $\tilde h^\pm$.}
\vspace*{-0.25cm}

\end{figure}

\begin{figure}[htbp]
\epsfig{file=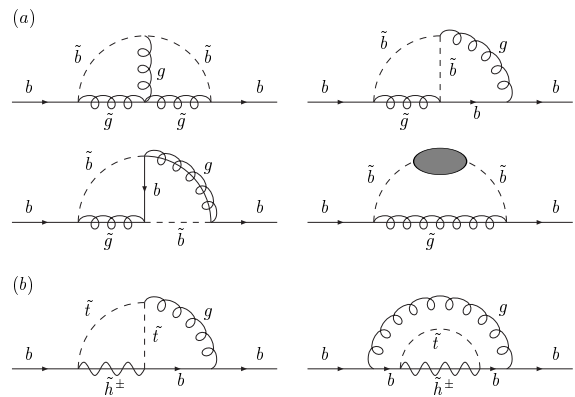,%
        bbllx=30pt,bblly=350pt,bburx=520pt,bbury=650pt,%
        scale=0.5,clip=}
\vspace*{-0.7cm}

\caption{\it \label{fg:two-loop} Typical two-loop diagrams of (a) the
SUSY-QCD and (b) the top-induced SUSY-electroweak contributions to the
bottom self-energy involving bottom quarks $b$, sbottom $\sq$ and stop
$\st$ squarks, gluons $g$, gluinos $\sgl$ and charged Higgsinos $\tilde
h^\pm$.}
\vspace*{-0.40cm}

\end{figure}

Typical SUSY-QCD and SUSY-electroweak diagrams that contribute to the
bottom self-energy at one-loop order are displayed in
Fig.~\ref{fg:one-loop}. The leading $\Dmb$ corrections can be obtained
from the diagrams in Fig.~\ref{fg:one-loop} with off-diagonal mass
insertions in the virtual sbottom and stop propagators in the chiral
squark basis \footnote{In this work we neglect the small contributions
of $A_b$ which have been included in the resummation in
Ref.~\cite{GHS}.}. These mass insertions yield a factor $\lambda_b \mu
v_2$ for the sbottom propagators and $\lambda_t A_t v_2$ in the stop
case. One obtains the one-loop results of Eq.~(\ref{eq:dmb}) by
replacing $v_2 \to \sqrt{2}\phi_2^{0*}$ and expressing the neutral Higgs
component $\phi_2^{0*}$ of the second Higgs doublet by the mass
eigenstates $h,H,A$ \cite{GHS}. These replacements lead to the exact
interactions with non-propagating Higgs fields, i.e.\,in the low-energy
limit of small Higgs momentum \cite{let}. This method will be applied to
the leading two-loop diagrams within SUSY--QCD. A typical sample of
two-loop diagrams contributing to the bottom self-energy is shown in
Figs.~\ref{fg:two-loop}a,b [a mass insertion has to be included in all
possible ways in the sbottom/stop propagators].

Dimensional regularization has been adopted for isolating the
ultraviolet singularities. The bottom momentum and its mass have been
put to zero while keeping the bottom Yukawa coupling $\lambda_b$ finite
in the mass insertions. All supersymmetric particles as well as the top
quark have been treated with full mass dependence. The two-loop vacuum
integrals have been reduced to the two-loop master integral
$T_{134}(m_1,m_3,m_4)$ \cite{t134} and one-loop one-point functions
$A_0(m)$ by standard reduction methods \cite{methods}. After
adding all two-loop diagrams linear ultraviolet divergences are left
over which are absorbed by the renormalization of all masses and
couplings appearing at one-loop order. The heavy masses $m_{\sq_i},
m_{\st_i}, m_{\sgl}$ appearing in the propagators have been
renormalized on-shell.  The trilinear coupling $A_t$ has been treated in
the on-shell scheme, too.  The strong coupling $\alpha_s$ and the top
Yukawa coupling $\lambda_t$ have been defined in the $\MS$ scheme with 5
active flavors, i.e.~the top quark and the supersymmetric particles
have been decoupled from the scale dependence of the strong coupling
$\alpha_s(\mu_R)$. Care has to be taken to include only the
desired order, i.e.~${\cal O}(\alpha_s^2 \mu \tgb/M_{SUSY})$ for
$\Dmb^{QCD}$ and ${\cal O}(\alpha_s \lambda^2_t A_t \tgb/M_{SUSY})$ for
$\Dmb^{elw}$. In this order the trilinear coupling $A_t$ only receives a
finite renormalization at NLO. More details of our calculation will be
published in \cite{preparation}.

Since dimensional regularization violates supersymmetry by
e.g.~attributing $(n-2)$ degrees of freedom to the gluons but two
degrees of freedom to its supersymmetric gluino partner, anomalous
counter terms have to be added in order to restore the supersymmetric
relations between the corresponding couplings. Since the strong coupling
factors at one-loop correspond to the Yukawa couplings between gluino,
sbottom and bottom quark states an anomalous counter term has to be
introduced in order to express all strong coupling factors in terms of
the conventional $\MS$ QCD coupling \cite{anom}. Moreover, the bottom
Yukawa coupling $\lambda_b$ in Fig.~\ref{fg:one-loop}a determines the
strength of the Higgs coupling to sbottom states at one-loop order which
differs from the $\MS$ bottom Yukawa coupling by a finite amount
\cite{anom}. The total anomalous counter term for $\Dmb^{QCD}$ is given
by [$C_A=3, C_F = 4/3$]
\bed
\delta \Delta m_{b;anom}^{QCD} = \left(\frac{C_A}{3}- \frac{C_F}{2}\right)
\frac{\alpha_s}{\pi} \Dmb^{QCD (1)}
\eed
For $\Dmb^{elw}$ the situation is similar. The one-loop order
involves the bottom and top Yukawa couplings of the Higgsino-bottom-stop
vertices as well as the top Yukawa coupling of the Higgs-stop-stop vertex.
Both are shifted from the $\MS$ couplings by finite amounts
\cite{anom}.  The sum of anomalous counter terms for $\Dmb^{elw}$ is
given by
\bed
\delta \Delta m_{b;anom}^{elw} = -C_F \frac{\alpha_s}{\pi} \Dmb^{elw (1)}
\eed
The final results have been included in the program HDECAY \cite{hdecay}
which calculates the masses and couplings of the MSSM Higgs bosons as
well as their decay widths and branching ratios.

The numerical analysis of the corrections $\Dmb$ and their impact on
neutral Higgs boson decays is performed for the ``small $\alpha_{eff}$''
MSSM scenario \cite{bench} as a representative case:
\bea
\tgb & = & 30,\;
M_{\tilde Q} = 800~{\rm GeV},\;
M_{\sgl} = 500~{\rm GeV} \\
M_2 & = & 500~{\rm GeV},\;
A_b = A_t = -1.133~{\rm TeV},\;
\mu = 2~{\rm TeV} \nonumber
\eea
For the Higgs masses and couplings \footnote{The mixing angles $\alpha,
\beta$ are derived from the RG-improved effective Higgs potential
consistently.} we use the RG-improved two-loop expressions of
Ref.\,\cite{rgi} The bottom quark pole mass has been chosen to be
$M_b=4.60$ GeV, which corresponds to a $\MS$ mass
$\overline{m}_b(\overline{m}_b)=4.26$~GeV.  The strong coupling constant
has been normalized to $\alpha_s(M_Z)=0.118$.
\begin{figure}[hbt]
\begin{picture}(200,360)(0,0)
\put(30,180){\epsfig{file=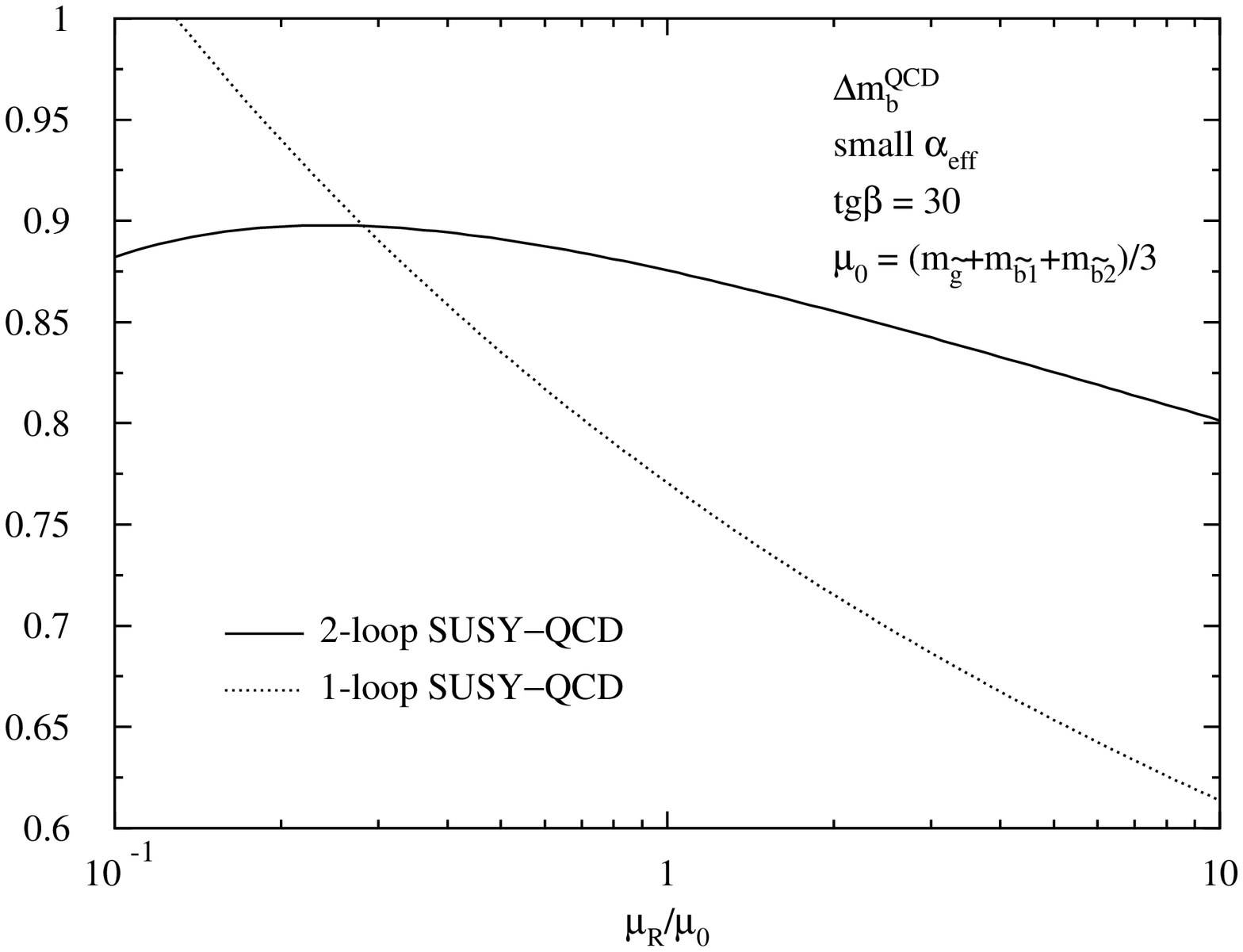,%
        bbllx=150pt,bblly=230pt,bburx=450pt,bbury=650pt,%
        scale=0.45,clip=}}
\put(30,0){\epsfig{file=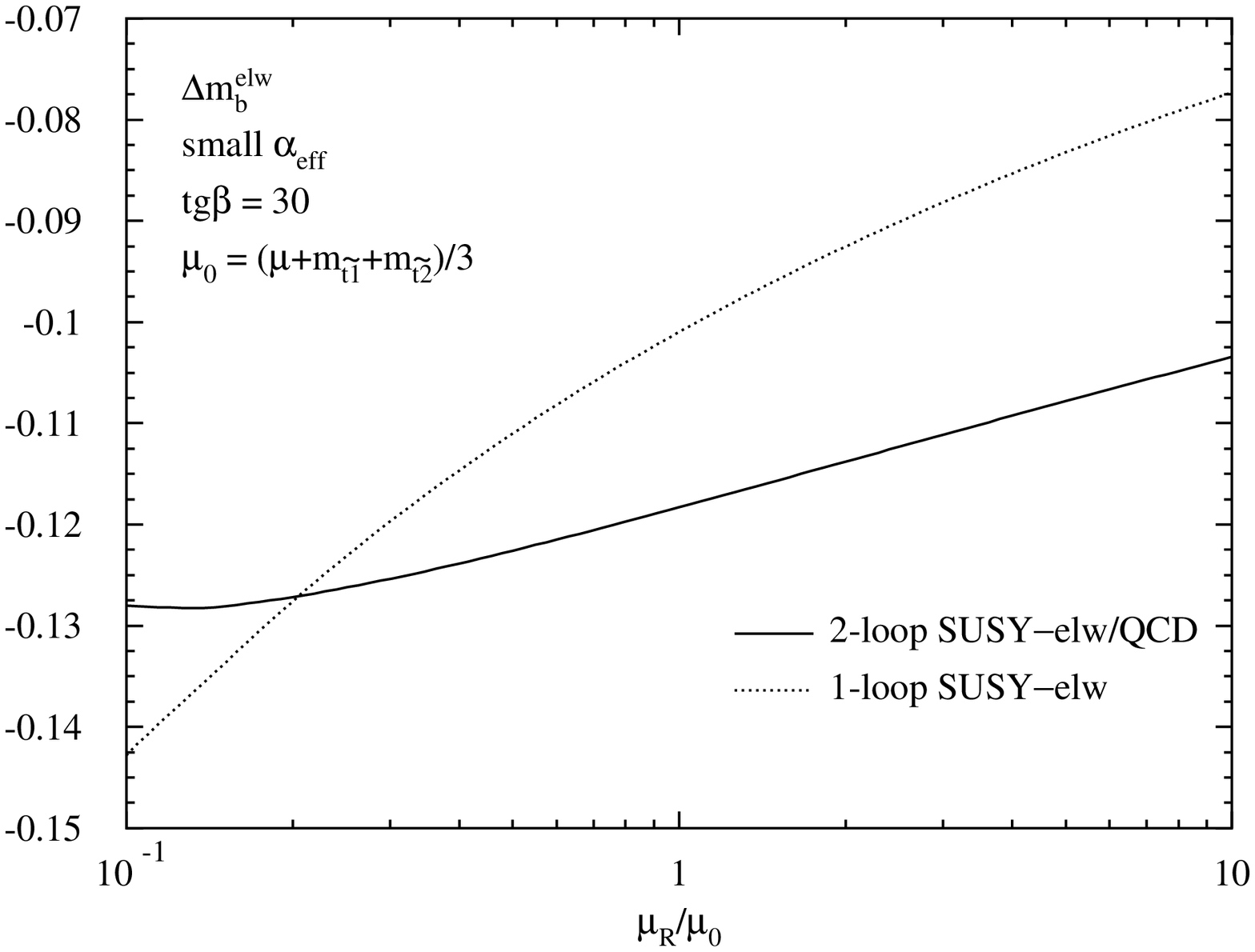,%
        bbllx=150pt,bblly=230pt,bburx=450pt,bbury=650pt,%
        scale=0.45,clip=}}
\put(100,340){$(a)$}
\put(100,160){$(b)$}
\end{picture}
\vspace*{-0.3cm}

\caption{\it \label{fg:dmb} Scale dependence of the corrections $\Dmb$
at one- and two-loop order: (a) the SUSY-QCD part $\Dmb^{QCD}$ and (b)
the SUSY-electroweak part $\Dmb^{elw}$ in the ``small $\alpha_{eff}$''
scenario.}
\vspace*{-0.60cm}

\end{figure}

The scale dependences of the corrections $\Dmb^{QCD}$ and $\Dmb^{elw}$
are displayed in Fig.~\ref{fg:dmb} at one- and two-loop order. The
central scale of the SUSY-QCD part $\Dmb^{QCD}$ is chosen as the average
of the SUSY-particle masses contributing at one loop,
i.e.~$\mu_0=(m_{\sq_1} + m_{\sq_2} + m_{\sgl})/3$, and as
$\mu_0=(m_{\st_1} + m_{\st_2} + \mu)/3$ for the SUSY-electroweak part
$\Dmb^{elw}$. We obtain a significant reduction of the scale dependence
at two-loop order and thus a large reduction of the theoretical
uncertainty. Moreover a broad maximum/minimum develops at scales of
about 1/4 to 1/3 of the chosen central scale in contrast to the
monotonous scale dependences at one-loop order. In the ``small
$\alpha_{eff}$'' scenario the SUSY-QCD corrections are large and
positive, while the SUSY-electroweak corrections are of moderate
negative size. However, the sign and size of the corrections depends on
the chosen MSSM scenario. The two-loop corrections amount to ${\cal
O}(10\%)$ in $\Dmb^{QCD}$ and a few per cent in $\Dmb^{elw}$ for the
central scale choices.
\begin{figure}[htbp]
\begin{picture}(200,520)(0,0)
\put(30,350){\epsfig{file=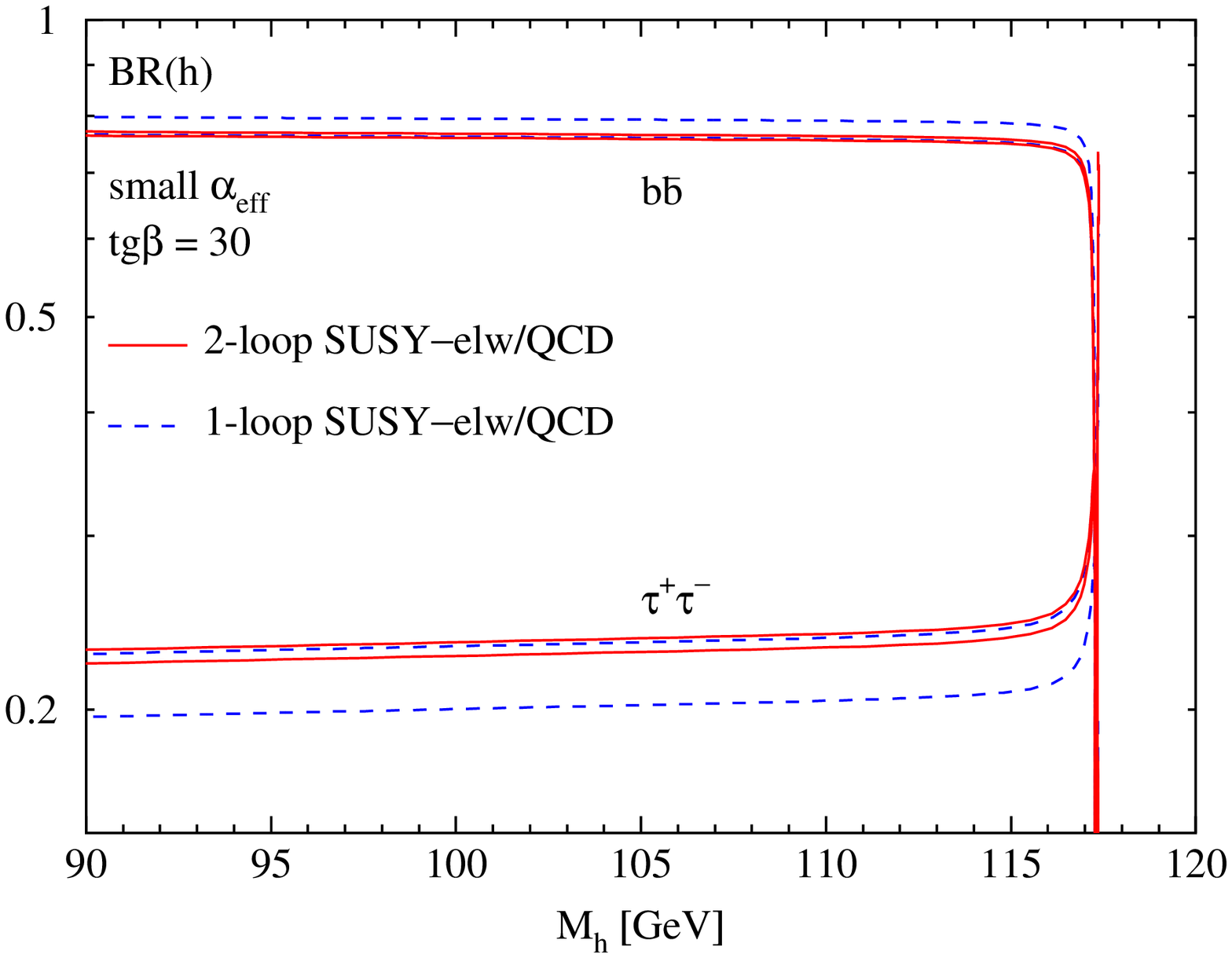,%
        bbllx=150pt,bblly=230pt,bburx=450pt,bbury=650pt,%
        scale=0.43,clip=}}
\put(30,175){\epsfig{file=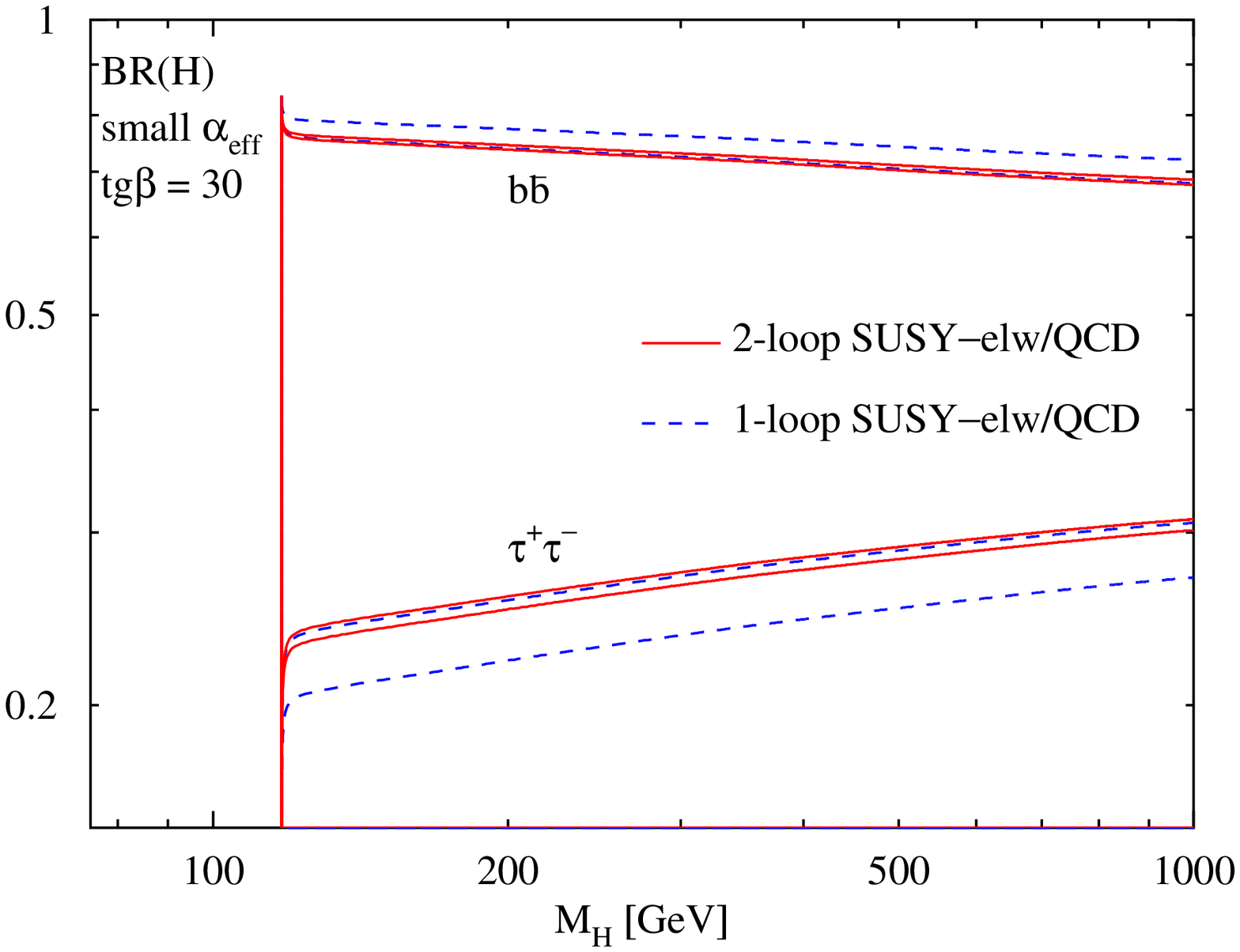,%
        bbllx=150pt,bblly=230pt,bburx=450pt,bbury=650pt,%
        scale=0.43,clip=}}
\put(30,0){\epsfig{file=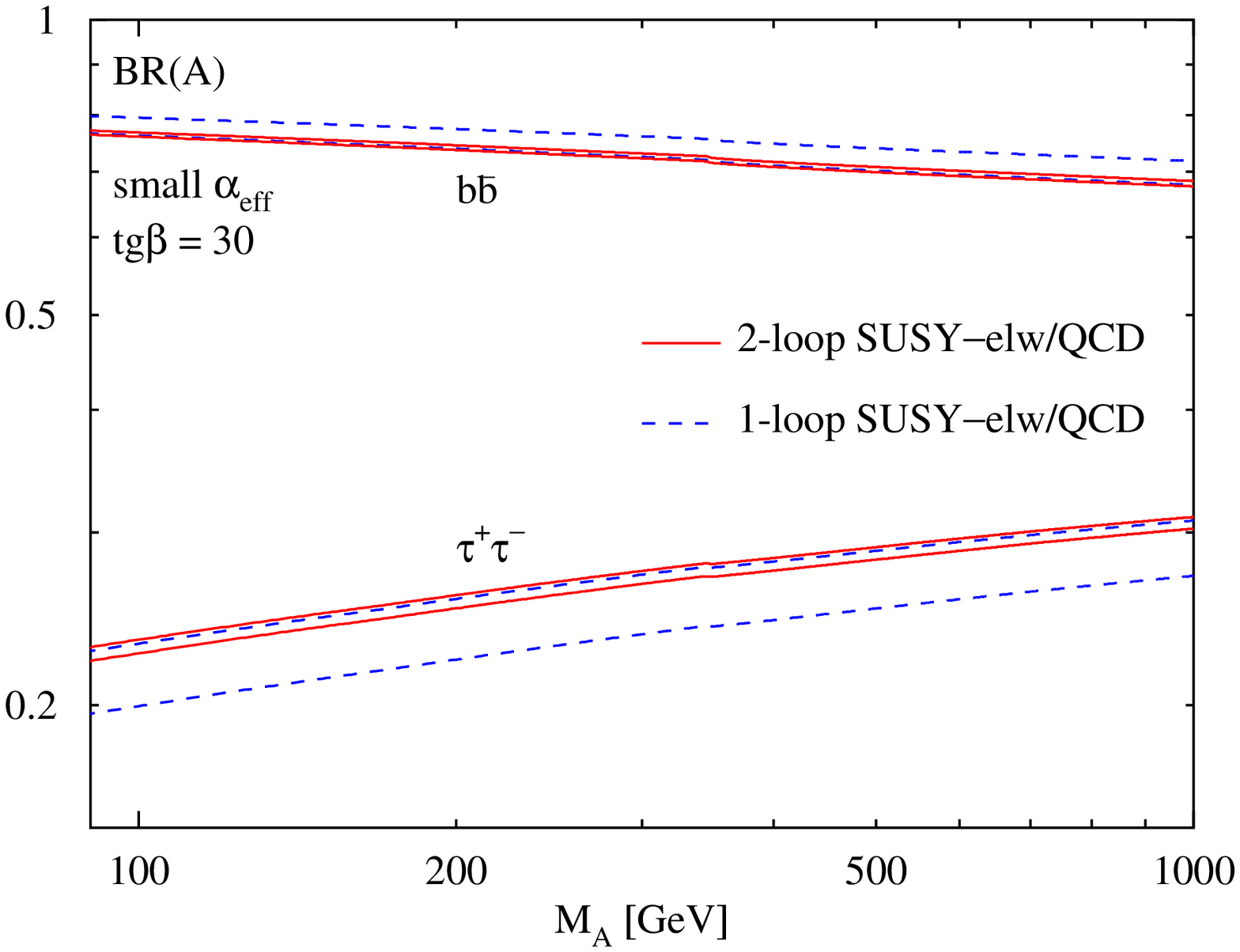,%
        bbllx=150pt,bblly=230pt,bburx=450pt,bbury=650pt,%
        scale=0.43,clip=}}
\put(185,500){$(a)$}
\put(185,325){$(b)$}
\put(185,150){$(c)$}
\end{picture}
\vspace*{-0.3cm}

\caption{\it \label{fg:brh} Branching ratios of (a) the light scalar,
(b) the heavy scalar and (c) the pseudoscalar Higgs boson in the ``small
$\alpha_{eff}$'' scenario. The dashed blue bands indicate the scale
dependence at one-loop order and the full red bands at two-loop order by
varying the renormalization scales between 1/3 and 3 times the central
scales given by the corresponding average SUSY particle masses.}
\vspace*{-0.70cm}

\end{figure}

The branching ratios of the neutral MSSM Higgs bosons are depicted in
Figs.~\ref{fg:brh}a-c. The bands at one-loop order (dashed blue curves)
and two-loop order (full red curves) are defined by varying the
renormalization scale between 1/3 and 3 times the corresponding central
scale of the SUSY-QCD and SUSY-electroweak parts. We only show the two
dominant decay modes into $b\bar b$ and $\tau^+\tau^-$ pairs.  The
uncertainties of the branching ratios reduce from ${\cal O}(10\%)$ at
one-loop order to the per-cent level at two-loop order. The per-cent
accuracy now matches the expected experimental accuracies at a future
linear $e^+e^-$ collider.

Since we have determined the effective resummed Yukawa coupling at
two-loop order the results will also affect all other processes which
are significantly induced by bottom Yukawa couplings, as e.g.~MSSM Higgs
radiation off bottom quarks at $e^+e^-$ colliders \cite{eebbh} and
hadron colliders \cite{ppbbh}. The two-loop corrections can easily be
included in the corresponding numerical programs.

In summary, the significant scale dependence of ${\cal O}(10\%)$ of the
NLO predictions for processes involving the bottom quark Yukawa
couplings of supersymmetric Higgs bosons requires the inclusion of NNLO
corrections. For the corrected Yukawa couplings, we find a reduction of
the scale dependence to the per-cent level at NNLO.  The improved NNLO
predictions for the bottom Yukawa couplings can thus be taken as a base
for experimental analyses at the Tevatron and the LHC as well as the
ILC.

\begin{acknowledgments}
The authors are pleased to thank J.~Guasch for many useful discussions
and carefully reading the manuscript. We are indebted to A.~Denner,
M.~M\"uhlleitner, H.~Rzehak and P.~Zerwas for comments on the
manuscript. This work is supported in part by the the Swiss National
Science Foundation and the European Community's Marie-Curie Research
Training Network HEPTOOLS under contract MRTN-CT-2006-035505.
\end{acknowledgments}

\end{document}